\documentclass[11pt,a4paper]{article}
\usepackage{amsmath,amssymb,mathrsfs}
\usepackage[dvipdfmx]{graphicx}

\newcommand{\ms}[1]{\mathscr{#1}}
\newcommand{\mc}[1]{\mathcal{#1}}

\def\un#1{\relax\ifmmode\@@underline#1\else
        $\@@underline{\hbox{#1}}$\relax\fi}

\let\du=\du                     % dot-under
\def\a{\alpha}
\def\b{\beta}

\def\d{\delta}

\def\f{\phi}
\def\g{\gamma}

\def\j{\psi}

\def\m{\mu}
\def\n{\nu}

\def\p{\pi}

\def\s{\sigma}

\def\L{\Lambda}

\def\ce{{\cal E}}

\def\cg{{\cal G}}

\def\car{{\cal R}}

\def\cw{{\cal W}}

\def\cy{{\cal Y}}

\def\bo{{\raise-.3ex\hbox{\large$\Box$}}}               % D'Alembertian
\def\pa{\partial}                                       % curly d
\def\TH{{\raise.2ex\hbox{$\displaystyle \bigodot$}\mskip-4.7mu \llap H \;}}
\def\face{{\raise.2ex\hbox{$\displaystyle \bigodot$}\mskip-2.2mu \llap {$\ddot
        \smile$}}}                                      % happy face
\def\Bar#1{\overline{#1}}                       % big bar
\def\VEV#1{\left\langle #1\right\rangle}        % < >
\def\leftrightarrowfill{$\mathsurround=0pt \mathord\leftarrow \mkern-6mu
        \cleaders\hbox{$\mkern-2mu \mathord- \mkern-2mu$}\hfill
        \mkern-6mu \mathord\rightarrow$}
\def\dvec#1{\vbox{\ialign{##\crcr
        \leftrightarrowfill\crcr\noalign{\kern-1pt\nointerlineskip}
        $\hfil\displaystyle{#1}\hfil$\crcr}}}           % <--> accent
\def\sfrac#1#2{{\vphantom1\smash{\lower.5ex\hbox{\small$#1$}}\over
        \vphantom1\smash{\raise.4ex\hbox{\small$#2$}}}} % alternate fraction
\def\bfrac#1#2{{\vphantom1\smash{\lower.5ex\hbox{$#1$}}\over
        \vphantom1\smash{\raise.3ex\hbox{$#2$}}}}       % "
\def\afrac#1#2{{\vphantom1\smash{\lower.5ex\hbox{$#1$}}\over#2}}    % "
\def\[{\lfloor{\hskip 0.35pt}\!\!\!\lceil}
\def\]{\rfloor{\hskip 0.35pt}\!\!\!\rceil}

\def\du#1#2{_{#1}{}^{#2}}

\def\ha{{\fracmm12}}

\def\un{\underline}
\def\fracmm#1#2{{{#1}\over{#2}}}

\def\low#1{{\raise -3pt\hbox{${\hskip 0.75pt}\!_{#1}$}}}

\newskip\humongous \humongous=0pt plus 1000pt minus 1000pt
\def\caja{\mathsurround=0pt}
\def\eqalign#1{\,\vcenter{\openup2\jot \caja
        \ialign{\strut \hfil$\displaystyle{##}$&$
        \displaystyle{{}##}$\hfil\crcr#1\crcr}}\,}
\newif\ifdtup

\newcommand{\be}{\begin{equation}}
\newcommand{\ee}{\end{equation}}
\newcommand{\nbe}{\begin{equation*}}
\newcommand{\nee}{\end{equation*}}

\newcommand{\lb}{\label}

\begin{document}

\begin{titlepage}

\begin{center}

April 2013 \hfill IPMU13-0076\\
\hfill ITP-UH -05/13\\
\hfill UT-13-11

\noindent
\vskip2.0cm
{\Large \bf 

New Actions for Modified Gravity and Supergravity

}

\vglue.3in

{\large
Sergei V. Ketov~${}^{a,b,c}$ and Takahiro Terada~${}^{d}$ 
}

\vglue.1in

{\em
${}^a$~Department of Physics, Tokyo Metropolitan University \\
Minami-ohsawa 1-1, Hachioji, Tokyo 192-0397, Japan \\
${}^b$~Kavli Institute for the Physics and Mathematics of the Universe (IPMU), \\
The University of Tokyo, 5-1-5 Kashiwanoha, Kashiwa, Chiba 277-8568, Japan \\
${}^c$~Institute for Theoretical Physics, Leibniz University of Hannover \\ 
Appelstrasse 2, 30167 Hannover, Germany\\
${}^d$~Department of Physics, The University of Tokyo,\\
7-3-1 Hongo, Bunkyo, Tokyo 113-0033, Japan
}

\vglue.1in
ketov@phys.se.tmu.ac.jp, takahiro@hep-th.phys.s.u-tokyo.ac.jp

\vglue.3in

\end{center}

\begin{abstract}
We extend the $f(R)$ gravity action by including a generic dependence upon the 
Weyl tensor, and further generalize it to supergravity by using the super-curvature ($\mc{R}$) and 
super-Weyl ($\mc{W}$) chiral superfields in $N=1$ chiral curved superspace. We argue that our
 (super)gravitational actions are the meaningful  extensions of the phenomenological $f(R)$ gravity 
and its locally supersymmetric  generalization towards their UV completion and their embedding into 
superstring theories. The proposed actions can be used for study of  cosmological perturbations and 
gravitational instabilities due to a nonvanishing Weyl tensor in gravity and supergravity.
\end{abstract}

\end{titlepage}

%%%%%%%%%%%%%%%%%%%%%%%%%%%%%%%%%%%%%%%%%%%%

\section{Introduction}\label{sec:Intro}

The $f(R)$ gravity theories, whose Lagrangian is given by the function $f$ of the 
spacetime scalar curvature $R$,
\be \lb{act}
S = -\ha \int d^4 x \, \sqrt{-g}\, f(R) ~~,
\ee
are the particular class of modified gravity theories which can provide the {\it geometrical} description 
of inflation in the early universe and acceleration of the present universe due to gravity alone --- see 
e.g., Refs.~\cite{tsu,ketrev} for a review --- in agreement with all known observations. 
The $f(R)$ gravity is known to be classically equivalent to the scalar-tensor gravity~\cite{oldr}, 
so that in the context of inflation or dark energy it amounts to 
quintessence. The ``fifth force'' at present due to exchange of the extra scalar (dubbed 
{\it scalaron} in the context of $f(R)$ gravity) can be effectively {\it screened} 
on local scales (like the Solar system) but can allow 
the enhancement of gravity on cosmological scales due to the so-called {\it chameleon} 
effect~\cite{chamel}. Gravitational instabilities in $f(R)$ gravity can also be avoided 
by demanding the proper signs of the first and second derivatives of the function $f$, 
thus making it free of ghosts and tachyons~\cite{ketrev}. The coupling of $f(R)$ gravity 
to matter fields after a transformation to the Einstein frame gives rise to the couplings 
of inflaton (scalaron) to all matter fields and thus leads to the {\it universal} reheating
 after inflation in the early universe~\cite{star}.
All the successes of the $f(R)$ gravity theory are 
related to the FLRW backgrounds.

The $f(R)$ gravity models currently have the {\it phenomenological} status, i.e. 
they are {\it not} (yet) derivable from any fundamental theory of gravity (like 
superstrings). Because of that any $f(R)$ gravity model needs {\it fine-tuning} of its 
parameters, in order to meet observations. 
Moreover, the $f(R)$ gravity is {\it neither} UV-complete {\it nor} renormalizable. The renormalizability
can be restored by adding the higher-curvature terms containing the Weyl tensor, like 
e.g., the conformal gravity term proportional to the Weyl tensor squared \cite{stelle}. 
Yet another way to improve the status of $f(R)$ gravity  is to find its embedding into the 
fundamental framework of superstring theory. It should be mentioned that the 
Weyl-tensor-dependent terms are known to appear in the (perturbative) superstring 
gravitational effective action indeed \cite{mybook}.  Hence, at the best, the $f(R)$ gravity 
may be considered as merely {\it part} of the gravitational effective action which is presumably 
derivable from a fundamental theory of quantum gravity (like superstrings). The $f(R)$ gravity  
part is responsible for the evolution of the scale factor in the FLRW metric of the universe, 
however, it is {\it not} enough for treating gravitational (tensor) perturbations. 
For example, the $f(R)$ gravity-based models of dark energy can only be distinguished from the 
standard ($\L$CDM)  Cosmological Model  by studying cosmological perturbations \cite{watak}.  
 Our interpretation makes it clear that the full gravitational 
action should have other terms beyond the $f(R)$ action. 

{}From this perspective it is natural to extend $f(R)$ gravity action (\ref{act}) to a more general one, namely,
\be \lb{baction}
S = -\ha\int d^4 x \sqrt{-g} f(R,C) ~~,
\ee
having a generic dependence upon the spacetime Weyl tensor $C_{\m\n\rho\sigma}=R_{\m\n\rho\sigma}
-\fracmm{1}{2}( g_{\m\rho}R_{\n\sigma}-g_{\n\rho}R_{\m\sigma} -g_{\m\sigma}R_{\n\rho}+ g_{\n\sigma}R_{\m\rho})+
\fracmm{1}{6}\left(  g_{\m\rho}g_{\n\sigma}- g_{\n\rho}g_{\m\sigma}\right)R$ also. 
Since the indices of the Weyl tensor have to be contracted in the Lagrangian, 
the $C$-dependence is actually given by
\be\lb{cexp}
\eqalign{
 f(R,C) & = f(R,C^2, C^3, C^4, \ldots) \cr
&  = f_0(R) + f_2(R)C^2 + f_3(R)C^3 + f_4(R)C^4 + {\cal O}(C^5) \cr}
\ee
where the $C^n$ denote the scalar products of the Weyl tensor, and the dots may also
include the contracted covariant derivatives of $R$ and $C$ as the additional arguments 
of the $f$-function. In the case of $f(R)$ gravity, adding the covariant derivatives of  $R$ 
leads to a classically equivalent scalar-tensor gravity with {\it more} scalars \cite{wands}.
In what follows we ignore the terms with the covariant derivatives of $R$ and $C$ 
for simplicity.
The FLRW background has
$ \lb{flrw} C^{\rm FLRW}_{\m\n\rho\sigma}=0 $
so that an arbitrary $C$-dependence in the action (\ref{baction}) does {\it not} affect the 
Friedman equation for the FLRW metric, and hence, keeps the cosmological achievements 
of $f(R)$ gravity. But, for example, the Schwarzschild solution and the black hole physics will be
modified \cite{kmo}.

When compared to a generic gravitational action, our action (\ref{baction}) is 
distinguished by the absence of manifest dependence upon the Ricci curvature tensor. At the 
{\it quadratic} level with respect to the curvatures its only possible contribution, which is 
proportional to the Ricci tensor squared, can always be eliminated via the Gauss-Bonnet 
(topological) combination in favor of the $C^2$ term. A generic dependence  of the gravitational 
action upon the Ricci tensor can lead to the extra propagating massless spin-2 mode \cite{maed}.

Our action (\ref{baction}) can also be  considered as the {\it alternative} to the popular $f(R,G)$ gravity 
where $G$ is the Gauss-Bonnet  combination, 
$G = C_{\m\n\rho\sigma} C^{\m\n\rho\sigma} -2R_{\m\n}R^{\m\n} +\frac{2}{3} R^2$.
The $G$-combination is a total derivative in four dimensions, so that the linear term in $G$ does not
affect the equations of motion, thus leading to a ghost-free $f(R,G)$ theory. The spectrum of the linearized 
$(R+C^2)$ action has a massive spin-2 ghost particle in addition to a massless graviton \cite{stelle}. Presumably,
this ghost violates unitarity in a quantized  $(R+C^2)$ field theory.~\footnote{See, however, Refs.~\cite{tomb,mann}
challenging the standard lore about non-unitarity of conformal gravity.} The unitarity issue is crucial for a
fundamental theory of gravity, but does not arise when treating the $C^2$ term as a perturbation in the action.
It may also be possible that the conformal gravity ghost is an artifact of the truncation of some highly non-linear
(with respect to the curvature) action to a four-derivative action. See Ref.~\cite{hassan} for the possible ghost-free 
completion of the conformal gravity by the partially massless bimetric gravity.

However, our main reasoning for the absence of the manifest Ricci tensor dependence in the
gravitational effective action is supersymmetry. We are going to demonstrate that our action 
(\ref{baction}) allows a locally $N=1$ supersymmetric extension as a  {\it chiral} supersymmetric 
invariant in curved superspace. Indeed, if such an action is to arise from superstrings, it must be in a 
supersymmetric context, while the chirality of the gravitational effective action would guarantee 
stability of its cosmological solutions against the higher-order quantum corrections due to the well 
known non-renormalization theorems in supersymmetry and supergravity \cite{a1,a2,a3,stelle2}.
As is well known in superspace supergravity \cite{a1,a2,a3,theisen}, the relevant superfield containing 
the  Ricci tensor as one of its field component in the superfield %supergravity
 is {\it not}
chiral, whereas the supergravity superfields containing the $R$ and $C$ tensors {\it are}
chiral (see Sec.~\ref{sec:superspace} below for more details). 

A supersymmetrization of Eq.~\eqref{baction} can also be considered as a supersymmetric generalization of 
the $F(\mc{R})$  supergravity action~\cite{gket} that is the manifestly $N=1$ supersymmetric extension of the 
$f(R)$ gravity action in $N=1$ chiral curved superspace,
\be \lb{faction}
 S_F = \int d^4xd^2\theta\,\ce F(\car) + {\rm H.c.}~~,
\ee
in terms of the analytic function $F(\car)$.~\footnote{The field construction of the $F(\mc{R})$ 
supergravity theory by using the  $N=1$ superconformal tensor calculus was given in Ref.~\cite{cec}.}
Besides having the manifest local $N=1$ supersymmetry, the action (\ref{faction}) has 
the so-called {\it auxiliary freedom} \cite{g12} because the auxiliary fields do 
{\it not} propagate in this theory. It distinguishes the action (\ref{faction}) from other possible 
supersymmetric extensions of Eq.~(\ref{act}). 
A calculation of the real function $f(R)$ in Eq.~(\ref{act}) from a given holomorphic 
function $F(\car)$ in Eq.~(\ref{faction}) requires solving an algebraic equation of motion for
the auxiliary field $M$. It is the non-trivial task in general, unlike the usual supergravity whose 
dependence upon the auxiliary fields is always Gaussian. The component structure of the 
bosonic sector of $F(\car)$ supergravity was systematically  investigated 
in Refs.~\cite{ket2,ket3,ket4,kwata,2kw} on the simplest examples.

Some physical applications of the $F(\car)$ supergravity theory to the early universe 
cosmology, inflation and reheating were systematically studied in Refs.~\cite{kstar,kstar2,ktsu,yoko}. 
In particular, a successful embedding of the chaotic slow roll (Starobinsky) inflation into the $F(\car)$ 
supergravity is  based on the following {\it Ansatz} \cite{kstar}:
\be \lb{fan}
F(\car) = -\fracmm{1}{2}f_1\car +\fracmm{1}{2}f_2\car^2-\fracmm{1}{6}f_3\car^3
\ee 
whose coefficients are given by 
\be \lb{fnota}
f_1 = \fracmm{3}{2} M^2_{\rm Pl}~, \qquad f_2=\sqrt{\fracmm{63}{8}} \fracmm{M^2_{\rm Pl}}{m}~,\quad  
{\rm and} \quad f_3=\fracmm{15M^2_{\rm Pl}}{M^2}
\ee
in terms of the scalaron masses: $M$ in the high curvature regime
and $m$ in the low curvature regime, respectively \cite{kstar,ktsu}. We have temporarily restored the 
Planck  mass dependence here, in order to show the (mass) dimensions of the $f$-coefficients.
A possible connection between the $F(\car)$ supergravity and the {\it Loop Quantum Gravity} 
was investigated in Ref.~\cite{gkn}. 

In this paper we generalize the $F(\mc{R})$ supergravity to a more general theory whose bosonic sector 
includes an $f(R,C)$ gravity action~\eqref{baction}.

Our paper is organized as follows. In Sec.~\ref{sec:Einstein} we rewrite the bosonic action (\ref{baction}) 
to the Einstein frame where the $R$-dependence is reduced to the standard Einstein-Hilbert 
term, in the presence of the propagating scalaron and the Weyl tensor. In Sec.~\ref{sec:superspace} we
construct a new manifestly supersymmetric extension of the bosonic action (\ref{baction}) by 
using curved superspace of the (old) minimal superspace supergravity. Sec.~\ref{sec:standard} is devoted to 
rewriting our new supergravity action to the more conventional form, in terms of the K\"ahler
potential and the ``superpotential". In Sec.~\ref{sec:bosonic} we derive the bosonic 
part of the simplest non-trivial model in our new family of modified supergravity theories. 
We discuss the possible origin of our new supergravity actions in Sec.~\ref{sec:discussion}. We use the natural  units $c=\hbar=M_{\rm Pl}=1$ where $M_{\rm Pl}$ 
is the  reduced Planck mass, and the $(1+3)$-dimensional space-time signature $(+,-,-,-)$.

\section{$f(R,C)$ gravity in Einstein frame} \label{sec:Einstein}

The action (\ref{baction}) is the extension of (\ref{act}) with an extra dependence upon the Weyl tensor. 
Hence, as long as the Weyl tensor vanishes, all the results of $f(R)$ gravity can be reproduced. 
For instance, the vacuum solutions in both theories with $R=R_0$ satisfy the equation
\be  \lb{vacs} R_0f'(R_0) =2 f(R_0).
\ee  

The generalized action (\ref{baction}) can be transformed to the Einstein frame, like
the $f(R)$ gravity action (\ref{act}). Let us rewrite the action (\ref{baction})  to the 
form
\be \lb{act1}
S =  -\ha\int d^4 x \, \sqrt{-g}\,\left[ f'(\f,C)(R-\f) +f(\f,C) \right]
\ee
where the new scalar field $\f$ has been introduced. The primes denote the derivatives
with respect to the {\it first} argument. On the one side, the equation of motion for the new scalar is algebraic,
\be \lb{eqmf}
 f''(\f,C)(R-\f)=0 .
\ee
Assuming that $f''\neq 0$, we get $\f=R$ and, hence, recover the original action 
(\ref{act}) back. 

On the other side, let us define a new metric
\be \lb{localw}
\tilde{g}_{\m\n}= f'(\f,C)g_{\m\n}
\ee
in the action (\ref{act1}), where the scalar function $f'$ is given by
\be \lb{deco}
f'(\f,C)= f'(\f,0) + \left. \fracmm{df'}{d(C^2)}\right|_{C=0} C^2 + {\cal O}(C^4) .
\ee
Though Eq.~(\ref{localw}) is not a standard  Weyl  transformation because the Weyl tensor $C=C(g)$ is metric-dependent, 
it can still be considered as the (non-canonical) local field redefinition of the metric, under which the Weyl tensor transforms {\it covariantly},
\be \lb{wtt}
\tilde{C}_{\m\n\rho\sigma}\equiv C_{\m\n\rho\sigma}(\tilde{g})= f'(\f,C)C_{\m\n\rho\sigma}(g) .
\ee 
As a result, the  action (\ref{act1}) takes the form
\be \lb{act2}
S =  \int d^4 x \, \sqrt{-\tilde{g}}\, \left[ -\fracmm{1}{2}\tilde{R} +
\fracmm{3}{4(f')^2}\tilde{g}^{\m\n}\pa_{\m}f'\pa_{\n}f' -V(\f,C) \right]
\ee
where we have introduced the scalar function
\be \lb{spot}
V (\f,C) =  \fracmm{  f(\f,C) - \f f'(\f,C) }{2f'(\f,C)^2} .
\ee
The new metric $\tilde{g}$ can be considered as the metric in the Einstein frame.
After the scalar field redefinition
\be \lb{cans}
\s = \sqrt{ \fracmm{3}{2} } \ln f'(\f,C) \quad {\rm or} \quad
f'(\f,C)= \exp\left [ \sqrt{\fracmm{2}{3} }\s \right]  ,
\ee
the scalar kinetic term in the action  (\ref{act2}) takes the {\it canonical} form,
and the action itself in terms of the new fields $\s$ and $\tilde{g}_{\m\n}$ 
reads
\be  \lb{act3}
S[\s,\tilde{g}] =  \int d^4 x \, \sqrt{-\tilde{g}}\, \left[ -\fracmm{1}{2}\tilde{R} +
\fracmm{1}{2}\tilde{g}^{\m\n}\pa_{\m}\s\pa_{\n}\s -V(\s,\tilde{C}) \right]
\ee
with the scalar function
\be \lb{spot2}
V (\s,\tilde{C}) = \frac{1}{2} e^{-2\sqrt{2/3}\s}f( \f(\s,\tilde{C}),e^{-\sqrt{2/3}\s}\tilde{C}) 
- \frac{1}{2} e^{-\sqrt{2/3}\s}\f(\s,\tilde{C})
\ee
where $\f(\s,\tilde{C})$ is the solution to the algebraic equation (\ref{cans}).

As a non-trivial simple example, let us consider the following action:
\be \lb{ex1}
f(R,C)= R-\fracmm{R^2}{6M^2} -b R C_{\m\n\rho\sigma}C^{\m\n\rho\sigma}
\ee
with the real parameter $b$. We find 
\be \lb{eder}
f'(\f,C)  =1-\fracmm{\f}{3M^2} -bC^2 =  e^{\sqrt{2/3}\s},
\ee
which can be easily solved for
\be \lb{ssol} 
\f = 3M^2 \left[ 1  -e^{\sqrt{2/3}\s}  -b e^{2\sqrt{2/3}\s}\tilde{C}^2 \right].
\ee
Hence, the transformed action in the Einstein frame takes the form
\be \lb{ex1ac}
S =  \int d^4 x \, \sqrt{-\tilde{g}}\, \left[ -\fracmm{1}{2}\tilde{R} + 
+\fracmm{1}{2}\tilde{g}^{\m\n}\pa_{\m}\s\pa_{\n}\s -V(\s,\tilde{C}^2) \right]
\ee
with the scalar function 
\be \lb{ex1sp}
V (\s,\tilde{C}^2) = \frac{3}{4}M^2\left (1- e^{-\sqrt{2/3}\s}
+b e^{\sqrt{2/3}\s}\tilde{C}^2 \right)^2 .
\ee
The inflaton scalar potential $V(\s)$ at $C=\tilde{C}=0$ is known to be quite suitable for slow roll inflation at large (positive) $\s$ \cite{llbook,kan}.  
However, a nonvanishing  Weyl tensor
in Eq.~(\ref{ex1sp}) may destabilize the slow-roll even at a small value of the parameter
$b$ due to the large exponential factor in front of the $\tilde{C}^2$ term.

As is clear from our derivation, the classically equivalent actions (\ref{baction})
and (\ref{act3}) are {\it not} equivalent in quantum theory because they are related
via the non-trivial field redefinition which results in the non-trivial field-dependent 
Jacobian in the path integral. 

Due to the presence of the $\tilde{C}^2$ term in the action (\ref{ex1ac}), this quantum
gravity theory is formally {\it renormalizable} but has {\it ghosts} in Minkowski background \cite{stelle}. In the context of a fundamental theory 
of quantum gravity any presence of ghosts is unacceptable \cite{kmy} so that the $C^2$ term may not be  allowed. However, in the perturbative 
framework,
when the gravity spectrum is determined by the leading (Einstein-Hilbert) action while all the other higher-derivative terms are considered  as the
interaction, the presence of the $\tilde{C}^2$ term is not a problem. For {\it linear} gravitational perturbations around the FLRW background, 
the whole $C$-dependence in the action (\ref{baction}) is irrelevant.

\section{$F(\car,\cw)$ supergravity in superspace}\label{sec:superspace}

In this Section we demonstrate that our action (\ref{baction}) has a simple chiral locally $N=1$ supersymmetric extension in 
four spacetime dimensions. For that purpose we use the chiral version of the curved superspace in the (old) minimal 
formulation of $N=1$ supergravity \cite{a1,a2,a3}. The curved  superspace is the most powerful, concise  and straightforward method 
of constructing general couplings in supergravity, in the manifestly 
supersymmetric way. We use the notation of Ref.~\cite{ketrev} and briefly comment on its relation to the more
standard notation of Ref.~\cite{a2} in Sec.~\ref{sec:bosonic}.

To reduce the off-shell field contents of superfield supergravity to the minimal set,  
one imposes certain off-shell constraints on the supertorsion tensor in curved 
superspace \cite{a1,a2,a3}. An off-shell supergravity multiplet has the auxiliary fields 
of noncanonical (mass) dimension, in addition to the physical spin-2 field (graviton) 
$e^a_{\m}$ and spin-3/2 field (gravitino) $\j_{\m}$. In the old minimal setting the auxiliary fields 
(in a WZ-type gauge) are given by a complex scalar $M$ and a real vector $b_{\m}$. 
It is worth mentioning  that imposing the off-shell constraints is independent upon writing a 
supergravity action.
 
The chiral superspace density reads
\be \lb{den}
\ce(x,\theta) = e \left[ 1 +i\theta\s^a\bar{\j}_a -\theta^2 
\left( M^* + \bar{\j}_a\bar{\s}^{ab}\bar{\j}_b\right) \right] \ee
where $e=\sqrt{-\det g}$, $\j^{\a}_a=e^{\m}_a\j_{\m}^{\a}$ is chiral gravitino, 
$M=S+iP$ is the complex scalar auxiliary field. We use the lower case middle Greek 
letters $\m,\n,\ldots=0,1,2,3$ for curved spacetime vector indices, the 
lower case early Latin letters $a,b,\ldots=0,1,2,3$ for flat (target) space 
vector indices, and the lower case early Greek letters $\a,\b,\ldots=1,2$ for 
chiral spinor indices.

A solution to the superspace Bianchi identities together with the constraints 
defining the $N=1$ Poincar\'e-type minimal supergravity theory reduce all the 
super-curvature and super-torsion tensor superfields to only {\it three} covariant 
tensor superfields, $\car$, $\cg_a$ and $\cw_{\a\b\g}$, subject to the off-shell 
relations \cite{a1,a2,a3}:
\be \lb{bi1}
 \cg_a=\bar{\cg}_a~,\qquad \cw_{\a\b\g}=\cw_{(\a\b\g)}~,\qquad
\bar{\ms{D}}_{\dot{\a}}\car=\bar{\ms{D}}_{\dot{\a}}\cw_{\a\b\g}=0~,\ee
and
\be \lb{bi2}
 \bar{\ms{D}}^{\dot{\a}}\cg_{\a\dot{\a}}=\ms{D}_{\a}\car~,\qquad
\ms{D}^{\g}\cw_{\a\b\g}=\frac{i}{2}\ms{D}\du{\a}{\dot{\a}}\cg_{\b\dot{\a}}+
\frac{i}{2}\ms{D}\du{\b}{\dot{\a}}\cg_{\a\dot{\a}}~~,\ee
where $(\ms{D}_{\a},\bar{\ms{D}}_{\dot{\a}},\ms{D}_{\a\dot{\a}})$ stand for the 
curved superspace $N=1$ supercovariant derivatives, and the bars denote 
Hermitian conjugation.

The covariantly chiral complex scalar superfield $\car$ has the scalar
curvature $R$ as the coefficient at its $\theta^2$ term, the real vector 
superfield $\cg_{\a\dot{\a}}$ has the traceless Ricci tensor, 
$R_{\m\n}+R_{\n\m}-\frac{1}{2}g_{\m\n}R$, as the coefficient at its 
$\theta\s^a\bar{\theta}$ term, whereas the covariantly chiral, complex, 
totally symmetric, fermionic superfield $\cw_{\a\b\g}$ has the self-dual part
of the Weyl tensor $C_{\mu\nu\rho\sigma}$ as the coefficient at its linear 
$\theta^{\d}$-dependent term. 

As is clear from Eqs.~(\ref{bi1}) and (\ref{bi2}), building a {\it chiral} superspace action 
(without using the covariant derivatives) is only possible with the superfields $\car$ 
and  $\cw_{\a\b\g}$. 

Hence, the $F(\car)$ supergravity action (\ref{faction}) admits a natural extension
in the chiral curved superspace because of the last equation (\ref{bi1}), namely,
\be \lb{waction}
 S = \int d^4xd^2\theta\,\ce F(\car,\cw) + {\rm H.c.}
\ee
with an extra dependence upon the totally symmetric spinor $N=1$ covariantly-chiral 
Weyl superfield  $\cw_{\a\b\g}$ of the old minimal $N=1$ superspace supergravity. Since
the $\cw_{\a\b\g}$ is anti-commuting and has only four independent components, an
expansion of the superfield $F(\car,\cw)$ in the $\cw_{\a\b\g}$ terminates as 
\be 
 F(\car,\cw)= F_0(\car) + F_2(\car) \cw^2 + F_4(\car)\cw^4 
\ee
in terms of the (complex) scalar products  of the Weyl superfield $\cw_{\a\b\g}$. 
For definiteness, we confine ourselves to the concrete supersymmetric model 
defined by \be \lb{mod}
 F(\car,\cw) = -\fracmm{1}{2}f_1\car +\fracmm{1}{2}f_2\car^2-\fracmm{1}{6}f_3\car^3
 + g \car \cw^2
\ee
with the real parameters $(f_1,f_2,f_3,g)$, which is the simplest $\cw$-dependent extension
of Eqs.~(\ref{fan}) and (\ref{fnota}). The (mass) dimension of the new coupling constant $g$ 
in Eq.~({\ref{mod}) is negative $(-1)$.

\section{K\"ahler potential and ``superpotential'' out of $F(\car,\cw)$}\label{sec:standard}

In this Section we show that the most general $F(\car,\cw)$ supergravity action (\ref{waction}) can be transformed 
in curved superspace (i.e. in the manifestly supersymmetric way) to the more conventional form, in terms of the K\"ahler potential 
and the ``superpotential''. After that going to the Einstein frame merely requires the standard procedure of Weyl transformations for 
the component fields \cite{a2}  or the super-Weyl transformations of the superfields \cite{Bagger:2000dh}. 

%%%%%   revised from here (v13)  %%%%%
 First, the action  (\ref{waction}) is classically equivalent to
\be \lb{efw}
S=\int d^4 x\,d^2 \theta\, {\cal E}\, \left[ -\cy \car +Z(\cy,\cw) \right] 
+{\rm H.c.}
\ee
where we have introduced the new (independent) covariantly chiral scalar superfield $\cy$ 
and the new analytic function  $Z(\cy,\cw)=\cy \car (\cy)  + F(\car (\cy) , \cw)$ as the Legendre transform of the function $F(\car , \cw)$ with respect to its first argument: the functional form of $\car (\cy)$ is the inverse of Eq.~\eqref{EoMofY}.  In fact, the equation of motion for $\cy$ is
\begin{align}
\cy = Z'^{-1}(\car ,\cw)=-F'(\car ,\cw) \label{EoMofY}
\end{align}
where derivatives (denoted by primes) and the inverse are with respect to the first arguments, considering the second argument $\cw$ as a parameter, and we assume $Z''(\cy ,\cw)\neq 0$ or equivalently $F''(\car ,\cw)\neq 0$.
 Substituting the solution $\cy(\car,\cw)$ back into the action (\ref{efw}) reproduces the original 
action (\ref{waction}).

%%%%%%%%%%   to here (v13)  %%%%%%%%%%

Now we treat $\cy$ as a dynamical superfield.
The kinetic terms of $\cy$ in the action (\ref{efw}) are obtained by using the (Siegel) identity 
\begin{align} \label{siegel}  
\int d^4 x\,d^2 \theta\, {\cal E}\, \cy \car  +{\rm H.c.}&=
\int d^4 x \, d^4 \theta \, E^{-1} (\cy +\bar{\cy} ) \nonumber \\
& = -\frac{3}{8} \int d^4 x \, d^2 \theta\, {\cal E}\, \left( \bar{\ms{D}}^2 -8\car \right) e^{-K/3} +{\rm H.c.}
\end{align}
where $E^{-1}$ is the full curved superspace density. and $K$ the K\"ahler 
potential of the superfields $(\cy,\Bar{\cy})$, 
\be \label{kahl}
K = -3 \ln \left( \cy + \bar{\cy} \right) + 3 \ln 3.
\ee
It gives rise to the "no-scale" kinetic terms
\be \label{kinc}
{\cal L}_{\rm kin} =  \left.
\fracmm{\pa^2 K}{\pa\cy \pa\bar{\cy}}\right|_{\cy=Y}
\pa_{\mu} Y\pa^{\mu}\bar{Y} = 3 \fracmm{ \pa_{\mu}Y\pa^{\mu}\bar{Y} } 
{(Y +\bar{Y})^2}.
\ee
These kinetic terms (\ref{kinc}) represent the {\it non-linear sigma model} 
\cite{mybook}  with the hyperbolic target space of (real) dimension two, 
whose metric is known as the standard (Poincar\'e) metric having the
 $SL(2,{\bf R})$  isometry.

Next, consider the remaining term $Z(\cy , \mc{W})$. For example, as regards our superfield Ansatz (\ref{mod}) with the notation (\ref{fnota}), we find
\be \lb{zan2} 
Z(\cy,\cw) =  \fracmm{7M^2}{40m^2} \car(\cy,\cw) + \left( \cy - \fracmm{3}{4} + g\cw^2\right)
\left[ \fracmm{\sqrt{14}M^2}{60m} + \fracmm{2}{3} \car(\cy,\cw) \right] 
\ee
where
\be \lb{zan1}
\car(\cy,\cw) = \fracmm{\sqrt{14}}{20} \fracmm{M^2}{m} \left[ 
1- \sqrt{  1+ \fracmm{80m^2}{21M^2}\left( \cy - \fracmm{3}{4} + g\cw^2\right)} \; \right] .
\ee
%where we have used the notation (\ref{fnota}).
As is clear from Eq.~(\ref{efw}), the holomorphic function $Z(\cy,\cw) $ plays the r\^ole of the {\it superpotential}. 
The truly scalar superpotential is given by $Z(\cy,0)$.

In conclusion, the $F(\mc{R},\mc{W})$ supergravity action can be rewritten to the form of the standard matter-coupled supergravity action --- see Eq.~\eqref{efw} ---
as a sum of Eqs.~\eqref{siegel} and \eqref{zan2}, in terms of the chiral scalar superfield $\cy$ and the chiral spinor superfield $\mc{W}$.

\section{Bosonic sector of $F(\car,\cw)$ supergravity}\label{sec:bosonic}

%For instance, 
The superfield action (\ref{waction}) of $F(\mc{R},\mc{W})$ supergravity leads to the following 
field theory action in terms of the superfield components:\footnote{
When comparing our notation to that of Ref.~\cite{a2}, one should take into account that in the latter 
the space-time signature is $(-,+,+,+)$, the normalization of chiral integration over the  
anti-commuting superspace coordinates $\Theta$'s differs from ours by the factor of $4$,
%(the same applies to the scalar products of the spinorial super-covariant derivatives)
 and the  definition of the Riemann
curvature differs by the sign.
}
\begin{align}
 \mathcal{L}= & \int d^{2} \theta \mc{E}  F\left( \mc{R}, \mc{W} \right) + \text{H.c.} \nonumber \\
=& -\left. \mc{E} \ms{D}\ms{D} F \right| - 2\left. \ms{D}^{\a}\mc{E}\ms{D}_{\a}F \right|
- \left. \ms{D}\ms{D} \mc{E} F \right| +\text{H.c.}
\end{align}
where the vertical bars stand for the lowest field components of each superfield in its expansion with respect to
the anti-commuting superspace coordinates. By using the results of Refs.~\cite{theisen,a2}  we find the bosonic part of the 
action above in the form
\begin{align}
\mathcal{L}_{\text{b}}=& -e \left( -\frac{1}{3}R+\frac{2}{3}ie_{a}{}^{\mu}\ms{D}_{\mu}b^{a}+\frac{4}{9}M^{*}M
-\frac{2}{9}b^{\mu}b_{\mu} \right) \left. \frac{\partial F}{\partial \mc{R}}\right | -4eM^{*}F| \nonumber \\
& -\frac{e\epsilon^{\eta \lambda}}{576}\left(  \sigma^{ab}_{\alpha\beta}\sigma^{cd}_{\gamma\lambda}C_{abcd}% - R_{\lambda \dot{\eta}\gamma}{}^{\dot{\eta}}{}_{\alpha\beta}
- i\epsilon_{\lambda \alpha}\sigma^{\mu}_{\beta\dot{\eta}}\ms{D}_{\mu}b_{\gamma}{}^{\dot{\eta}}\right) \left( \sigma^{ef}_{\delta\epsilon}\sigma^{gh}_{\zeta\eta}C_{efgh}%-R_{\eta\dot{\kappa}\delta}{}^{\dot{\kappa}}{}_{\epsilon\zeta}
-i\epsilon_{\eta\delta}\sigma^{\nu}_{\epsilon\dot{\kappa}}\ms{D}_{\nu}
b_{\zeta}{}^{\dot{\kappa}}\right)  \left. \frac{\partial^{2}F}{\partial \mc{W}_{\delta\epsilon\zeta}\partial \mc{W}_{\alpha\beta\gamma}}\right| \nonumber \\
& +\text{H.c.}  \label{L_boson}
\end{align}
where all the fermionic contributions are ignored. The necessary formulae needed to derive Eq.~(\ref{L_boson})  are collected in Appendix~\ref{sec:app}.
 The equation of motion for the auxiliary complex scalar field $M$ reads
\begin{align}
0=&\frac{\partial \mc{L}_{\text{b}}}{\partial M^{*}} \nonumber \\
=
& -e F| +\frac{e}{6}M \left. \frac{\partial F^{\dag}}{\partial \mc{R}^{\dag}}\right | -\frac{e}{9}M\left( \left. \frac{\partial F}{\partial \mc{R}} \right | + \left. \frac{\partial F^{\dag}}{\partial \mc{R}^{\dag}} \right | \right) \nonumber \\
&  +\frac{e}{36}\left( -\frac{1}{3}R+\frac{2}{3}ie_{a}{}^{\mu}\ms{D}_{\mu}b^{a}+\frac{4}{9}M^{*}M-\frac{2}{9}b^{\mu}b_{\mu} \right) \left. \frac{\partial^{2} F^{\dag}}{\partial \mc{R}^{\dag 2}}\right | \nonumber \\
&-\frac{e \epsilon^{\dot{\eta} \dot{\lambda}} }{13824}\left( \bar{\sigma}^{ab}_{\dot{\alpha}\dot{\beta}}\bar{\sigma}^{cd}_{\dot{\gamma}\dot{\lambda}}C_{abcd} %R_{\eta \dot{\lambda}}{}^{\eta}{}_{\dot{\gamma}\dot{\alpha}\dot{\beta}}
+i\epsilon_{\dot{\lambda} \dot{\alpha}}\sigma^{\mu}_{\eta\dot{\beta}}\ms{D}_{\mu}b^{\eta}{}_{\dot{\gamma}}\right) \left( \bar{\sigma}^{ef}_{\dot{\delta}\dot{\epsilon}}\bar{\sigma}^{gh}_{\dot{\zeta}\dot{\eta}}C_{efgh}%R_{\kappa\dot{\eta}}{}^{\kappa}{}_{\dot{\delta}\dot{\epsilon}\dot{\zeta}}
 +i\epsilon_{\dot{\eta}\dot{\delta}}
\sigma^{\nu}_{\kappa\dot{\epsilon}}\ms{D}_{\nu}b^{\kappa}{}_{\dot{\zeta}} \right)  \left. \frac{\partial^{3}F^{\dag}}{\partial \bar{\mc{W}}_{\dot{\delta}\dot{\epsilon}\dot{\zeta}}\partial \bar{\mc{W}}_{\dot{\alpha}\dot{\beta}\dot{\gamma}} \partial \mc{R}^{\dag}}\right| . \label{EoMofM}
\end{align}

As regards our model (\ref{mod}), Eq.~\eqref{EoMofM} takes the form
\begin{multline}
 f_{3}M^{3}+3f_{3}MM^{*2}+6f_{2}M^{2}+12f_{2}MM^{*}-72f_{1}M %\nonumber
 \\
-f_{3}\left( 2 b_{\mu}b^{\mu}+6i e_{a}{}^{\mu}\ms{D}_{\mu}b^{a}+3R \right) M^{*}  -6f_{2}\left( 2 b_{\mu}b^{\mu}+6i e_{a}{}^{\mu}\ms{D}_{\mu}b^{a}+3R \right) %\nonumber
 \\
 -\frac{27}{4}g \left[ 2\left( C_{\mu\nu\rho\sigma}C^{\mu\nu\rho\sigma}-iC_{\mu\nu\rho\sigma}\tilde{C}^{\mu\nu\rho\sigma} \right) -\frac{4}{3}\left( F_{\mu\nu}F^{\mu\nu}+i F_{\mu\nu}\tilde{F}^{\mu\nu} \right)\right]=0
\end{multline}
where $F_{\mu\nu}=\partial_{\mu}b_{\nu}-\partial_{\nu}b_{\mu}$ is the field strength of the auxiliary vector field $b_{\mu}$, $\tilde{F}^{\mu\nu}=\frac{1}{2}\epsilon^{\mu\nu\rho\sigma}F_{\rho\sigma}$ is its Poincar\'e dual, and $\tilde{C}^{\mu\nu\rho\sigma}=\frac{1}{2}\epsilon^{\mu\nu\xi\pi}C_{\xi\pi}{}^{\rho\sigma}$. For many physical applications (as well as initial study) the imaginary  parts of the scalar fields may be ignored, so that $M$ is real. Then the above equation is simplified to
\begin{multline} \lb{cubeq}
f_{3}M^{3}+\frac{9}{2}f_{2}M^{2}-\left( \frac{3}{4}f_{3}R +\frac{1}{2}f_{3}b_{\mu}b^{\mu}+18f_{1} \right)M %\nonumber
 \\
-\frac{9}{2}f_{2}R-3f_{2}b_{\mu}b^{\mu}-\frac{27}{4}g \left( C_{\mu\nu\rho\sigma}C^{\mu\nu\rho\sigma}-\frac{2}{3}F_{\mu\nu}F^{\mu\nu} \right)=0 \text{.}
\end{multline}
It is always possible to get the real roots of this cubic equation by demanding positivity of its discriminant (e.g., via the standard Cardano-Vieta method) 
in the case of a sufficiently high scalar curvature and a sufficiently small contribution of the last term  in Eq.~(\ref{cubeq}). However, the corresponding formulae
 appear to be long and not very illuminating. Here we confine ourselves  to the much  simpler case when $f_{2}=f_{3}=0$. Then Eq.~\eqref{EoMofM}  can be easily 
solved as
\begin{align}
M=-\frac{g}{16 f_{1}}\left[ 3 \left(  C_{\mu\nu\rho\sigma}C^{\mu\nu\rho\sigma}-iC_{\mu\nu\rho\sigma}\tilde{C}^{\mu\nu\rho\sigma} \right) -2\left( F_{\mu\nu}F^{\mu\nu}+i F_{\mu\nu}\tilde{F}^{\mu\nu} \right) \right] .
\end{align}
Substituting the result with $f_{1}=3/2$ to the bosonic part of the Lagrangian~\eqref{L_boson}, we find 
\begin{align} \lb{finex}
e^{-1}\mc{L}_{\text{b}}=& -\frac{1}{2}R-\frac{1}{3}b_{\mu}b^{\mu}+\frac{g^{2}}{432}\left[ \frac{9}{4}\left( C_{\mu\nu\rho\sigma}C^{\mu\nu\rho\sigma} \right)^2 -\frac{9}{2} C_{\mu\nu\rho\sigma}C^{\rho\sigma\zeta\pi}C_{\zeta\pi\tau\phi}C^{\tau\phi\mu\nu} \right. \nonumber \\ 
&  +9 C_{\mu\nu\rho\sigma}C^{\rho\sigma}{}_{\zeta\pi}C^{\mu \zeta}{}_{\tau\phi}C^{\tau\phi\nu \pi}  -3C^{\mu\nu\rho\sigma}C_{\mu\nu\rho\sigma}F^{\zeta\pi}F_{\zeta\pi} 
-6 F^{\mu\nu}C_{\mu\nu\rho\sigma}C^{\rho\sigma\zeta\pi}F_{\zeta\pi} \nonumber \\
& \left. +12F^{\mu \zeta}F^{\nu \pi}C_{\mu\nu\rho\sigma}C_{\zeta\pi}{}^{\rho\sigma}-\left( F^{\mu\nu}F_{\mu\nu}\right)^{2} + 4 F^{\mu\nu}F_{\nu\rho}F^{\rho\sigma}F_{\sigma\mu}   \right] \text{.}
\end{align}

\section{Discussion}\label{sec:discussion}

In this Section we briefly comment on the possible {\it origin} of the Weyl-tensor-dependent terms in the gravitational 
effective action.

The obvious source of those terms is provided by the standard {\it Weyl anomaly} in the quantum field theory  of massless matter 
in a gravitational background, which is given by
\be \lb{wano}
\VEV{T}_g = \fracmm{c}{16\p^2} C^2 - \fracmm{a}{16\p^2}G
\ee
where we have introduced the trace $T$ of the matter energy-momentum tensor in the gravitational background $(g)$,
the central charge $c$, the Gauss-Bonnet combination $G$, and the so-called $a$-coefficient. Equation (\ref{wano}) 
suggests  us to consider even more general actions of the type
\be  \lb{mgen}
 S = -\ha\int d^4 x \sqrt{-g} f(R,C,G).
\ee
However, as was already mentioned above, the truly chiral supersymmetric extension is not compatible with the $G$-dependence. It implies that 
any dynamics (or gravity models) relying on the $G$-dependence of the gravitational  action may be unstable, being not protected against 
quantum corrections by supersymmetry. 

The supergravity corrections  in the $\alpha'$-expansion of the gravitational superstring effective action arise as the loop corrections in the quantized supergravity, though with {\it finite} coefficients.  Since the supergravity counterterms are 
usually given by the {\it full} superspace integrals, it is unlikely that  our action (\ref{waction}) can be generated in the 
{\it perturbative} superstring theory. However, it may well be generated {\it non-perturbatively}. The kinetic terms of the 
dilaton-axion (complex) scalar in superstring theory  are precisely given by the non-linear sigma-model (\ref{kinc}) indeed, whereas  the dilaton-axion superpotential can only be generated non-perturbatively in superstring theory.

\section*{Acknowledgements}

This work was supported by the World Premier International Research Center Initiative (WPI Initiative), MEXT, Japan. SVK was supported in part by the Tokyo Metropolitan University and the German Academic Exchange Service (DAAD). TT was supported by the grant of the Advanced Leading Graduate Course for Photon Science at the University of Tokyo.

\appendix
\section{The lowest components of the superfields}\label{sec:app}
The leading (in the zeroth order with respect to the Grassmann superspace coordinates) field components of various superfields can be obtained by using Refs.~\cite{theisen, a2}.  
Here we list the leading terms of the relevant bosonic superfields used in the main text:
\begin{align}
\mc{E}|=&e ,\\
%\ms{D}\ms{E}|=&ie\sigma^{a}\bar{\psi}_{a} \\
\ms{D}\ms{D}\mc{E}|=&4eM^{*}+4e\bar{\psi}_{\mu}\bar{\sigma}^{\mu\nu}\bar{\psi}_{\nu} ,\\
\mc{R}|=&-\frac{1}{6}M ,\\
%\bar{\mathscr{D}}_{\dot{\alpha}}R^{\dag}|=& -\frac{1}{6}\left(\bar{\psi}_{ab}\bar{\sigma}^{b}\sigma^{a}-ib^{a}\bar{\psi}_{a}+i\psi_{a}\sigma^{a}M^{*}  \right)_{\dot{\alpha}}\\
\ms{D}_{\alpha}\ms{D}_{\beta}\mc{R}|=& \frac{1}{2}\epsilon_{\alpha\beta} \left( -\frac{1}{3}R-\frac{2}{3}i\bar{\psi}^{\mu}\bar{\sigma}^{\nu}\psi_{\mu\nu}-\frac{1}{12}\epsilon^{\mu\nu\rho\sigma}\left( \bar{\psi}_{\mu}\bar{\sigma}_{\nu}\psi_{\rho\sigma}+\psi_{\mu}\sigma_{\nu}\bar{\psi}_{\rho\sigma} \right) \right. \nonumber \\
&\left. +\frac{2}{3}ie_{a}{}^{\mu}\ms{D}_{\mu}b^{a}+\frac{4}{9}M^{*}M-\frac{2}{9}b_{\mu}b^{\mu}-\frac{1}{3}\bar{\psi}^{\mu}\bar{\psi}_{\mu} M +\frac{1}{3}\psi_{\mu}\sigma^{\mu}\bar{\psi}_{\nu}b^{\nu} \right) , \displaybreak[2]\\
\ms{D}_{\delta}\mc{W}_{\alpha\beta\gamma}|=& \frac{1}{8}\left [ -2\sigma^{ab}_{\alpha\beta}\sigma^{cd}_{\gamma\delta} C_{abcd}  -i\psi_{\gamma\dot{\epsilon}\delta}\bar{\psi}_{\beta\dot{\eta}\alpha}{}^{\dot{\eta}\dot{\epsilon}}-i\bar{\psi}_{\gamma\dot{\epsilon}}{}^{\dot{\epsilon}}\psi_{\delta\dot{\eta}\alpha}{}^{\dot{\eta}}{}_{\beta}-\psi_{\alpha\dot{\epsilon}\beta}\bar{\psi}_{\gamma}{}^{\dot{\epsilon}}{}_{\dot{\eta}}b_{\delta}{}^{\dot{\eta}}  -\psi_{\gamma\dot{\epsilon}\delta}\bar{\psi}_{\alpha\dot{\eta}}{}^{\dot{\eta}}b_{\beta}{}^{\dot{\epsilon}} \right ]_{\text{t.s.}}\nonumber \\
&-\frac{i}{4}\epsilon_{\delta(\alpha}\hat{D}_{\beta}{}^{\dot{\epsilon}}b_{\gamma )\dot{\epsilon}} , \lb{last}
\end{align}
where we have introduced the notation 
\begin{align}
\hat{D}_{\beta \dot{\epsilon}}b_{\gamma \dot{\eta}}=&\ms{D}_{\beta\dot{\epsilon}}b_{\gamma\dot{\eta}} \nonumber \\
& +\frac{3}{2}\psi_{\beta\dot{\epsilon}}{}^{\alpha}\left[ \frac{1}{4}\bar{\psi}_{\gamma}{}^{\dot{\beta}}{}_{\alpha\dot{\beta}\dot{\eta}}+ \frac{1}{12}\epsilon_{\alpha\gamma}\bar{\psi}^{\beta\dot{\beta}}{}_{\beta\dot{\eta}\dot{\beta}}-\frac{i}{6}\psi_{\gamma\dot{\eta}\alpha}M^{*} +\frac{i}{6}\bar{\psi}_{(\gamma\dot{\beta}}{}^{\dot{\beta}}b_{\alpha )\dot{\eta}}-\frac{i}{12}\bar{\psi}_{\alpha}{}^{\dot{\beta}}{}_{\dot{\eta}}b_{\gamma\dot{\beta}}  \right] \nonumber \\
& -\frac{3}{2}\bar{\psi}_{\beta\dot{\epsilon}}{}^{\dot{\alpha}} \left[ \frac{1}{4}\psi^{\beta}{}_{\dot{\alpha}\beta\dot{\eta}\gamma}+ \frac{1}{12}\epsilon_{\dot{\alpha}\dot{\eta}}\psi_{\gamma}{}^{\dot{\beta}\beta}{}_{\dot{\beta}\beta}+\frac{i}{6}\bar{\psi}_{\gamma\dot{\eta}\dot{\alpha}} M -\frac{i}{6}\psi_{\beta ( \dot{\alpha}}{}^{\beta}b_{\gamma \dot{\eta})} +\frac{i}{12}\psi^{\beta}{}_{\dot{\alpha}\gamma}b_{\beta\dot{\eta}} \right] ~.
\end{align}
The subscript t.s. in Eq.~(\ref{last}) 
 denotes the total symmetrization of the undotted indices inside the square brackets, and $\psi_{\mu\nu}=\ms{D}_{\mu}\psi_{\nu}-\ms{D}_{\nu}\psi_{\mu}$.  The Hermitian vector field $b_{a}$ is defined as the lowest component of the superfield $\mc{G}_{a}$, $\mc{G}_{a}|=-\frac{1}{3}b_{a}$.

\end{document}